\documentstyle[preprint,prl,aps,epsf]{revtex}
\begin{document}
\title{Higher Order Force Gradient Symplectic Algorithms }
\author{Siu A. Chin and Donald W. Kidwell\footnote[1]{Present address:
Candescent Technologies,
6580 Via Del Oro,
San Jose, CA  95119
}}
\address{Center for Theoretical Physics, Department of Physics, 
Texas A\&M University, 
College Station, TX 77843}
\date{\today}
\maketitle
\begin{abstract}

We show that a recently discovered fourth order symplectic algorithm, which
requires one evaluation of force gradient in addition to three evaluations 
of the force, when iterated to higher order, yielded algorithms 
that are far superior to similarly iterated higher order algorithms
based on the standard Forest-Ruth algorithm. 
We gauge the accuracy of each algorithm by comparing the step-size 
independent error functions associated with energy conservation and 
the rotation of the Laplace-Runge-Lenz vector when solving a highly 
eccentric Kepler problem. For orders 6, 8, 10 and 12, the 
new algorithms are approximately a factor of $10^3$, $10^4$, 
$10^4$ and $10^5$ better.
     
\end{abstract}
\pacs{PACS:  }
\section {Introduction}
%\narrowtext

Symplectic algorithms\cite{serna,yoshi} for solving classical dynamical problems exactly
conserve all Poincar$\acute{\rm e}$ invariants. For periodic orbits, the errors in
energy conservation are bounded and periodic. This is in sharp contrast to 
Runge-Kutta type algorithms, whose energy error increases linearly with
integration time, even for periodic orbits\cite{shita,gladman}. Thus, symplectic 
algorithms are ideal for long time integration of equations of motion in problems 
of astrophysical interest\cite{wisdom}. For long time integrations, higher order 
algorithms are desirable because they permit the use of larger time steps. Symplectic 
algorithms are also advantageous in that higher order algorithms can be systematically 
generated from any low, even order algorithm\cite{creutz,su,yoshid}. 
In this work, we will show that higher order algorithms 
generated by a fourth order force gradient symplectic algorithm\cite{chin}, 
have energy errors that are several orders of magnitude smaller than 
existing symplectic algorithms of the same order. For completeness, we will 
briefly summarize the operator derivation of symplectic algorithms and their 
higher order construction in Section II. While the materials in this section 
are not new, we believe that we have restated Creutz' and Gockach's\cite{creutz} 
triplet construction of higher order algorithms in its most transparent setting. 
In sections III and IV we recall force gradient algorithms and discuss two 
distinct ways of gauging the errors of an algorithm when solving the Kepler 
problem. We present our results and conclusions in Sections V and VI.

\section {Operator Factorization and Higher Order Constructions}
After a tortuous start\cite{rutho,forest}, symplectic algorithms can be derived most simply 
on the basis of operator decomposition or factorization. 
For any dynamical variable $W(q_i,p_i)$, its time evolution is given by the
Poisson bracket,
\begin{equation}
{{d }\over{dt}}W(q_i,p_i)=\{W,H\}\equiv
          \sum_i\Bigl(
		          {{\partial W}\over{\partial q_i}}
                  {{\partial H}\over{\partial p_i}}
				 -{{\partial W}\over{\partial p_i}}
                  {{\partial H}\over{\partial q_i}}
				                    \Bigr).
\label{peq}
\end{equation}
If the Hamiltonian is of the form, 
\begin{equation}
H(p,q)={1\over{2}}\sum_ip_i^2+V(\{q_i\}),
\label{ham}
\end{equation}
the evolution equation (\ref{peq}) can be written as an operator
equation
\begin{equation}
{{dW }\over{dt}}=\sum_i\Bigl(
   {p_i}{{\partial }\over{\partial q_i}}
              +F_i{{\partial }\over{\partial p_i}}
                  \Bigr)W,
\label{peqop}
\end{equation}
with formal solution
\begin{equation}
W(t)={\rm e}^{t(T+V)}W(0),
\label{peqform}
\end{equation}
where $T$ and $V$ are first order differential operators defined by
\begin{equation}
T\equiv\sum_i
   {p_i}{{\partial }\over{\partial q_i}},\qquad
V\equiv\sum_i F_i{{\partial }\over{\partial p_i}}.
\label{tandv}
\end{equation}
Their exponentiations, ${\rm e}^{\epsilon T}$ and ${\rm e}^{\epsilon V}$, are
displacement operators which displace $q_i$ and $p_i$ forward in time via  
\begin{equation}
q_i\rightarrow q_i+\epsilon{p_i}$\qquad${\rm and}\qquad 
p_i\rightarrow p_i+\epsilon F_i.
\label{peqv}
\end{equation}
Thus, if ${\rm e}^{\epsilon (T+V)}$ can be factorized into products of 
the displacement operators ${\rm e}^{\epsilon T}$ and ${\rm e}^{\epsilon V}$,
each such factorization gives rise to an algorithm for 
evolving the system forward in time. For example, the second order
factorization
\begin{equation}
{\cal T}_{2}(\epsilon)\equiv
{\rm e}^{{1\over 2}\epsilon T}{\rm e}^{\epsilon V}
{\rm e}^{{1\over 2}\epsilon T}
={\rm exp}[\epsilon (T+V)+\epsilon^3C++O(\epsilon^5)\cdots],
\label{tb}
\end{equation}
corresponds to the second order algorithm
\begin{eqnarray}
{\bf q}_1=&&{\bf q}_0+{1\over 2}\epsilon{\bf p}_0,\nonumber\\
{\bf p}_1=&&{\bf p}_0+\epsilon{\bf F}({\bf q}_1),\nonumber\\
{\bf q}_2=&&{\bf q}_1+{1\over 2}\epsilon{\bf p}_1,
\label{altbs}
\end{eqnarray}
where ${\bf q}_0$, ${\bf p}_0$ and ${\bf q}_2$, ${\bf p}_1$ are the 
initial and final states of the algorithm respectively. This second 
order symplectic algorithm only requires one evaluation of the force.

The bilaterally symmetric form of ${\cal T}_{2}(\epsilon)$ automatically
guarantees that it is {\it time-reversible},
\begin{equation}
{\cal T}_{2}(-\epsilon){\cal T}_{2}(\epsilon)=1,
\label{timer}
\end{equation}
and implies that $\log({\cal T}_{2})$ can only be an odd function of 
$\epsilon$, as indicated in (\ref{tb}). The explicit form of the 
operator $C$ is not needed for our present discussion.

Consider now the the symmetric triple product
\begin{equation}
{\cal T}_{2}(\delta)
{\cal T}_{2}(-s\delta)
{\cal T}_{2}(\delta)
={\rm exp}[(2-s)\delta(T+V)+(2-s^3)\delta^3C
+O(\delta^5)+\cdots].
\label{tripd}
\end{equation}
This algorithm evolves the system forward 
for time $\delta$, backward for time
$s\delta$ and forward
again for time $\delta$. Since it is manifestly 
time-reversible, its error terms must be odd powers of $\delta$ only. 
Morever, its leading first and third order terms can only be the
sum of the first and third order terms of each constituent algorithm
as indicated. This is because non-additive terms must come from commutators
of operators and the lowest order non-vanishing commutator has to
have two first order terms and one third order term, which is fifth order.
The form of (\ref{tripd}) naturally suggests that the third order error term
can be made to vanish by choosing
\begin{equation}
s=2\,^{1/3}.
\end{equation}
Thus if we now rescale $\delta$ 
back to the standard step size by setting $\epsilon=(2-s)\delta$, the 
resulting triplet product would be correct to 4th order,
\begin{equation}
{\cal T}_{4}\equiv
{\cal T}_{2}\Bigl({\epsilon\over{2-s}}\Bigr)
{\cal T}_{2}\Bigl({-s\, \epsilon\over{2-s}}\Bigr)
{\cal T}_{2}\Bigl({\epsilon\over{2-s}}\Bigr)
={\rm exp}[\,\epsilon (T+V)
+O(\epsilon^5)+\cdots].
\label{tren}
\end{equation}
Expanding out the ${\cal T}_{2}\,$'s gives the explicit form:
\begin{equation}
{\cal T}_{4}\equiv
{\rm e}^{a_1\epsilon T}{\rm e}^{b_1\epsilon V}
{\rm e}^{a_2\epsilon T}{\rm e}^{b_2\epsilon V}
{\rm e}^{a_2\epsilon T}{\rm e}^{b_1\epsilon V}
{\rm e}^{a_1\epsilon T}
\label{texp}
\end{equation}
where, by inspection
\begin{equation}
a_1={1\over 2}{1\over{2-s}},\quad a_2=-{1\over 2}{{s-1}\over{2-s}},
\quad b_1={1\over{2-s}},\quad {\rm and}
\quad b_2=-{s\over{2-s}}.
\label{tcof}
\end{equation}
This fourth order symplectic algorithm was apparently obtained
by E. Forest in 1987. However, its original derivation was very complicated
and was not published with Ruth\cite{forest} until 1990. During this period
many groups, including Campostrini and Rossi\cite{camp} in 1990, 
Candy and Rozmous\cite{candy}
in 1991, independently published the same algorithm. Our discussion followed 
the earliest published derivation of this algorithm by Creutz and 
Gocksch\cite{creutz}
in 1989. After they were informed of this algorithm by Campostrini, they
provided the triplet construction and generalized it to higher order. 
The triplet construction was also independently published by Suzuki\cite{su}
and Yoshida\cite{yoshid} in 1990. 

Higher order algorithms can be obtained by repeating this construction.
Starting with any $n$th order symmetric algorithm, 
\begin{equation}
{\cal T}_{n}(\epsilon)
={\rm exp}[\,\epsilon (T+V)
+\epsilon^{(n+1)}D+\cdots],
\label{ford}
\end{equation}
the triplet product
\begin{equation}
{\cal T}_{n}(\delta)
{\cal T}_{n}(-s\delta)
{\cal T}_{n}(\delta)
={\rm exp}[(2-s)\delta(T+V)+(2-s^{n+1})\delta^{n+1}D
+O(\delta^{n+3})+\cdots],
\label{tpdford}
\end{equation}
will be of order $(n+2)$ if we choose 
\begin{equation}
s=2\,^{1/(n+1)}
\end{equation}
and renormalize $\delta=\epsilon/(2-s)$ as before. 

\section {Force Gradient Algorithms}

     The method of operator factorization can be applied to
many different classes of evolution equations. However, the triplet 
concatenations with a negative time step are a special construction 
with more limited applicability. For example, one cannot use it to derive 
similar Diffusion Monte Carlo or finite temperature path 
integral algorithms, because one cannot simulate diffusion backward in time
nor sample configurations with negative temperatures. The triplet construction
is a special example of Suzuki's\cite{nogo} general proof that, beyond second order, 
it is impossible to factorize ${\rm e}^{\epsilon (T+V)}$ only 
into products of ${\rm e}^{\epsilon T}\,$'s and 
${\rm e}^{\epsilon V}\,$'s without introducing negative time steps. 
For symplectic
algorithms this means that one can never develop a purely positive time step
fourth order algorithm by evaluating only the force. For many years 
the Forest-Ruth (FR) algorithm was the only known fourth order symplectic 
algorithm. Recently, a deeper understanding of the operator factorization 
process has yielded three new symplectic algorithms\cite{chin} all with purely
positive time steps. 
These new algorithms circumvented Suzuki's no-go theorem
by evaluating the force and its gradient. This
corresponds to factorizing ${\rm e}^{\epsilon(T+V)}$ in terms of operators
$T$, $V$, and the commutator $[V,[T,V]]$. The latter corresponds to
\begin{equation}
[V,[T,V]]=2F_j{ {\partial  F_i}\over{\partial q_j} }
                { {\partial  }\over{\partial p_i} }=
                \nabla_i|{\bf F}|^2
                { {\partial  }\over{\partial p_i} },
\end{equation}
which is the gradient of the squared magnitude of the force.
Of the three algorithms derived by Chin\cite{chin}, algorithm C is particularly 
outstanding and corresponds to the factorization
\begin{equation}
{\rm e}^{\epsilon(T+V)}= 
  {\rm e}^{\epsilon {1\over 6} T} 
  {\rm e}^{\epsilon {3\over 8} V}
  {\rm e}^{\epsilon {1\over 3} T} 
  {\rm e}^{\epsilon {1\over 4} \widetilde V}
  {\rm e}^{\epsilon {1\over 3} T} 
  {\rm e}^{\epsilon {3\over 8} V}
  {\rm e}^{\epsilon {1\over 6} T} 
+ O(\epsilon^5).
\label{ruth4}
\end{equation}
 where
\begin{equation}
\widetilde V=V+{1\over 48}\epsilon^2[V,[T,V]]. 
\label{ruthv4}
\end{equation}
The algorithm itself can be read off directly as
\begin{eqnarray}
{\bf q}_1&&={\bf q}_0+{1\over 6}\epsilon {\bf p}_0,\nonumber\\
{\bf p}_1&&={\bf p}_0+{3\over 8}\epsilon {\bf F}({\bf q}_1)\nonumber\\
{\bf q}_2&&={\bf q}_1+{1\over 3}\epsilon {\bf p}_1\nonumber\\
{\bf p}_2&&={\bf p}_1+{1\over 4}\epsilon \Bigl[{\bf F}({\bf q}_2)+
{1\over {48}}\epsilon^2\nabla|{\bf F}({\bf q}_2)|^2\Bigr]\\
\label{algc}
{\bf q}_3&&={\bf q}_2+{1\over 3}\epsilon {\bf p}_2\nonumber\\
{\bf p}_3&&={\bf p}_2+{3\over 8}\epsilon {\bf F}({\bf q}_3)\nonumber\\
{\bf q}_4&&={\bf q}_3+{1\over 6}\epsilon {\bf p}_3.\nonumber
\end{eqnarray}

In \cite{chin} it was shown that the maximum energy error for this
algorithm, when used to solve Kepler's problem, is smaller than
that of the FR algorithm by a factor of 80.
At the moment there is no general method for constructing higher order 
algorithms with only positive time steps. It is not even known 
whether a positive time step 6th order algorithm exists. Thus, beyond
4th order the triplet contruction is still the only systematic way of
generating higher order algorithms. In this work we show that intrinsic
error functions associated with higher order
algorithms generated from Chin's algorithm C are far smaller than those
generated from the FR algorithm.

\section {The Energy and the LRL Vector}

We gauge the numerical effectiveness of each algorithm by solving 
the two dimensional Kepler problem
\begin{equation}
{{d^2{\bf q}}\over{dt^2}}=-{ {\bf q}\over{q^3} },
\end{equation}
with initial conditions ${\bf q}_0=(10,0)$ and ${\bf p}_0=(0,1/10)$.
The resulting highly eccentric (e=0.9) orbit provides a non-trivial
testing ground for trajectory integration. 

A symmetric $n$th order symplectic 
algorithm evolves this system forward in time with Hamiltonian
\begin{equation}
H({\bf p},{\bf q})
=H_0({\bf p},{\bf q})+\epsilon^n H_n({\bf p},{\bf q})+O(\epsilon^{n+2}),
\label{hal}
\end{equation}
which deviates from the exact Hamiltonian 
$H_0({\bf p},{\bf q})={1\over 2}{\bf p}^2-1/|{\bf q}|$ 
by an error term $\epsilon^n H_n({\bf p},{\bf q})$ as indicated. 
To gauge the intrinsic merit of each algorithm we compare their step-size 
independent error coefficient $H_n({\bf p},{\bf q})$. This can be 
extracted numerically as follows. Let's start the system with total energy
$E_0=H_0({\bf p}(0),{\bf q}(0))$. Since the Hamiltonian (\ref{hal}) 
is conserved by the algorithm, we have   
\begin{equation}
E_0=H_0({\bf p}(t),{\bf q}(t))+\epsilon^n H_n({\bf p}(t),{\bf q}(t))
+O(\epsilon^{n+2})
\label{halc}
\end{equation}
Denoting $E(t)\equiv H_0({\bf p}(t),{\bf q}(t))$ and
 $H_n(t)=H_n({\bf p}(t),{\bf q}(t))$, we therefore have
\begin{equation}
H_n(t)=-\lim_{\epsilon\rightarrow 0}
{1\over {\epsilon^n}}\Bigl[ E(t)-E_0\Bigr ].
\end{equation}

    Energy conservation does not directly measure how well the orbit 
is determined. When the time step is not too small, a very noticeable
error is that the orbit precesses. One can, but it
is tedious, directly monitor this orbital precession\cite{gladman}. 
It is more expedient to follow the rotation of the Laplace-Runge-Lenz (LRL)
vector:
\begin{equation}
{\bf A}={\bf p}\times{\bf L}-\hat{\bf r}.
\end{equation}
When the orbit is exact the LRL vector is constant, pointing along the
semi-major axis of the orbit. When the orbit precesses the LRL vector
rotates correspondingly. 

For an $n$th order algorithm, 
\begin{equation}
{{d{\bf A} }\over{dt}}=\epsilon^n\sum_i\Bigl(
		          {{\partial {\bf A}}\over{\partial q_i}}
                  {{\partial H_n}\over{\partial p_i}}
				 -{{\partial {\bf A}}\over{\partial p_i}}
                  {{\partial H_n}\over{\partial q_i}}
				                    \Bigr).
\label{aeq}
\end{equation}
Thus, the rate of change of each component of the LRL vector is of order
$\epsilon^n$. The components themselves, which are time integrals of the 
above modulo a constant term, must also be of order $\epsilon^n$. 
Let the LRL vector initially be of length ${A_0}$ and lie along the
x-axis, then we have
\begin{eqnarray}
A_x(t)&&=A_0+\epsilon^n A_{nx}(t)+O(\epsilon^{n+2}),\\
A_y(t)&&=\epsilon^n A_{ny}(t)+O(\epsilon^{n+2}).
\end{eqnarray}
Since the square of the LRL vector is related to the energy by
\begin{equation}
{\bf A}^2=2{\bf L}^2E+1,
\end{equation}
the longitudinal deviation coefficient $A_{nx}(t)$ is related to the
energy error coefficient by
\begin{equation}
A_{nx}(t)=
{1\over {\epsilon^n}}{{\bf L}^2\over{A_0}}(E(t)-E_0)=-{{\bf L}^2\over{A_0}}H_n(t),
\end{equation}
which gives no new information. The perpendicular deviation coefficient 
$A_{ny}(t)$ is best measured in terms of the rotation angle: 
\begin{equation}
\theta(t)=\tan^{-1}\biggl[{{A_y(t)}\over{A_x(t)}}\biggr]=
\epsilon^n\biggl[{{A_{ny}(t)}\over{A_0}}\biggr]+\cdots
\end{equation}
To compare algorithms we again extract and compare their rotation error 
coefficient function $\theta_n(t)=A_{ny}(t)/A_0$ 
via
\begin{equation}
\theta_n(t)=\lim_{\epsilon\rightarrow 0}
{1\over {\epsilon^n}}\theta(t).
\end{equation}
Since this rotational angle is related to some integral of the energy 
error function, it is a better measure of the overall error of the algorithm.

\section {Results of Comparing Higher Order Algorithms}

 	By use of the triplet construction, we generated 6th, 8th, 10th, and
12th order algorithms from both the Forest-Ruth and Chin's C algorithm.
We computed the fractional energy deviation, which is just the 
negative of the energy error coefficient normalized by the 
initial energy,
\begin{equation}
\lim_{\epsilon\rightarrow 0}
{1\over {\epsilon^n}}\Bigl[ {{E(t)}\over{E_0}}-1\Bigr ]=-{{H_n(t)}\over{E_0}}.
\end{equation}
Smaller and smaller time steps $\epsilon$ are used until the extracted
coefficient function is stablized independent of the time step size. 
This typically occurs in the neighborhood of $\epsilon=P/5000$, where P is the 
period of the orbit.

\begin{figure}%%%%%%%%%%%%%%%%%%%%%%%%%%%%%%%%%%%%%%%%%%%%%%%%%%%%%%%%%
%\topinsert
\noindent
\vglue 0.2truein
\hbox{
\vbox{\hsize=7truein
\epsfxsize=6truein
\leftline{\epsffile{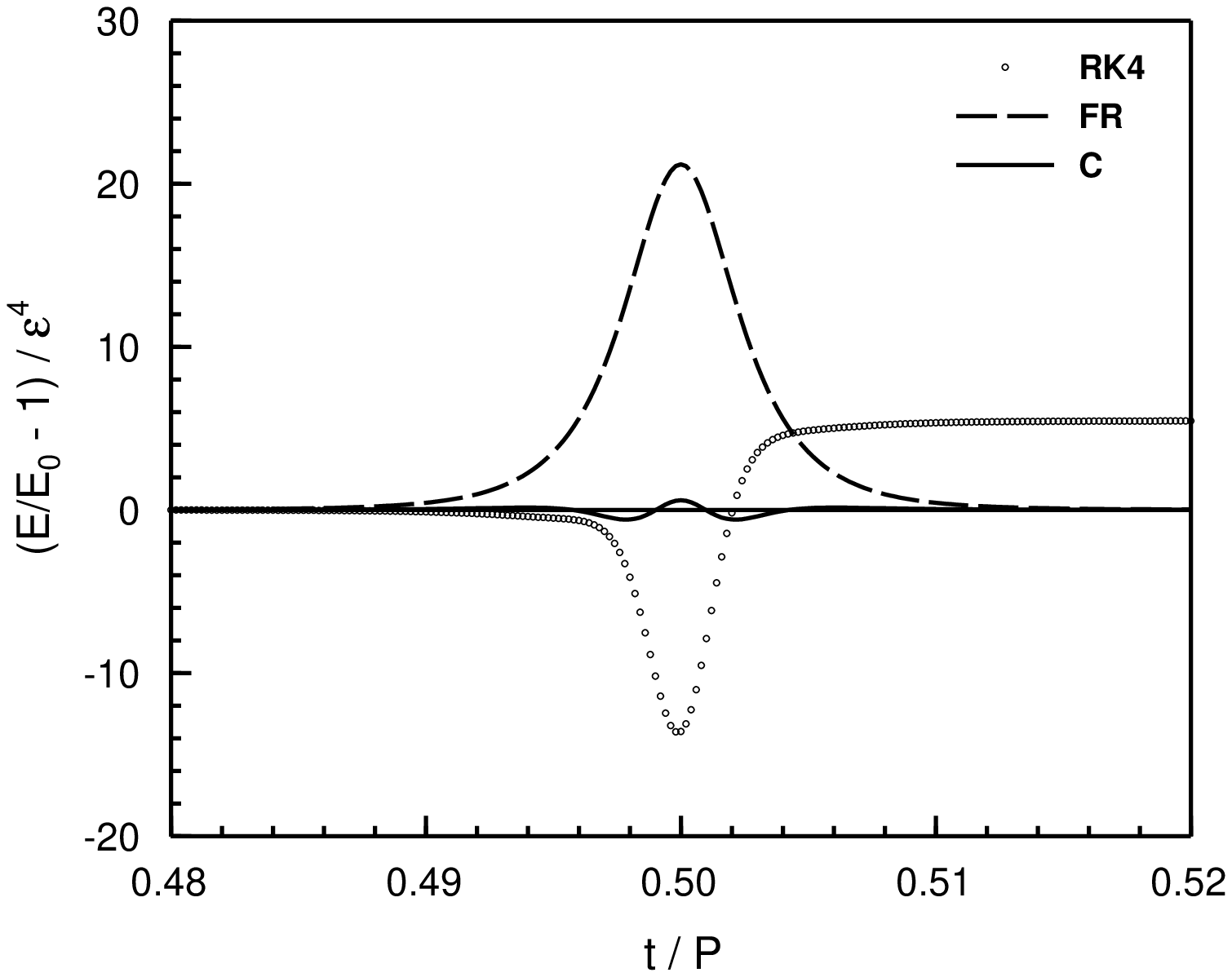}}
}
}
\caption{The normalized energy deviation of a particle in a Keplerian 
orbit, which measures the step-size
independent energy error coefficient $-H_4({\bf p}(t),{\bf q}(t))/E_0$. 
P is the period of the elliptical orbit and $\epsilon$ is the time step size. 
RK4, FR, and C denote results for the 4th order Runge-Kutta, Forest-Ruth, 
and Chin's C algorithm respectively. The maximum deviations for algorithm
FR and C are 21 and 0.27 respectively.
}
\label{fone}
\end{figure}

Fig.\ref{fone} compares the (negative) normalized 
error coefficient functions for the 4th order Runge-Kutta, Forest-Ruth and 
Chin's C algorithms over one period of the orbit. The error function for the
two symplectic algorithms are substantial only near mid period when the
particle is at its closest approach to the attractive center. For 
symplectic algorithms energy is conserved over one period, or its
non-conservation is periodic. Its average energy error is bounded and 
constant as a function of time. In contrast, the 4th order Runge-Kutta
energy error function is an irreversible, step-like function over one period.
Each successive period will increase the error by the same amount resulting
in a linearly rising, staircase-like error function in time. 
As noted earlier, the maximum error in Chin's algorithm C is smaller than that of 
the FR algorithm by a factor of 80. However, this error height comparison 
at one point is not meaningful. It is better to compare the energy error averaged 
over one period. This would require the integral of the energy error function. 
On this basis algorithm C will be better still. While the energy error 
integral can be done, the same goal can be achieved by 
monitoring the rotation of the LRL vector.      
    
\begin{figure}%%%%%%%%%%%%%%%%%%%%%%%%%%%%%%%%%%%%%%%%%%%%%%%%%%%%%%%%%
%\topinsert
\noindent
\vglue 0.2truein
\hbox{
\vbox{\hsize=7truein
\epsfxsize=6truein
\leftline{\epsffile{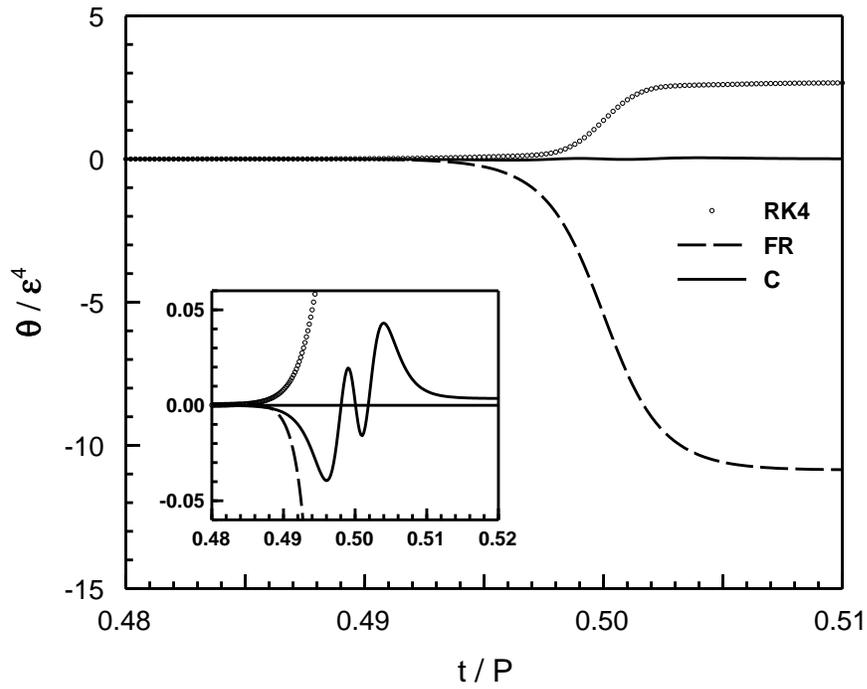}}
}
}
\caption{The step-size independent error coefficient of the rotation
angle of the Laplace-Runge-Lenz vector for 4th order algorithms. 
The LRL vector rotates substantial 
only when the particle is near mid period, closest to the attractive center.
The insert make visible the fine structure produced by algorithm C.
}
\label{ftwo}
\end{figure}

Fig.\ref{ftwo} shows the corresponding error coefficient functions of the
rotational angle of the LRL vector. After each period, the algorithms rotate the 
LRL vector by a definite amount. The error coefficient provides 
an intrinsic, step-size independent way of comparing this rotation. In  
Fig.\ref{ftwo} the rotated angle produced by algorithm C is too small 
to be visible when plotted on the same scale as the other algorithms. The
insert gives an enlargement of the details. The rotational angle 
of the LRL vector appears to be related to some integral of energy error function. 
Although we have not been able to demonstrate this analytically, numerical
integration of the energy error function does give a function similiar in
shape to the angle coefficient function, having the same numbers of maxima and
minima. For the Runge-Kutta, Forest-Ruth and Chin's C 4th order algorithms, the 
magnitudes of this rotation cofficient after one period are 2.666, 10.860, 
and 0.004 respectively. On this basis, algorithm C is better than FR by 
a factor of $\approx 3000$. When the orbit is integrated over 
many periods the rotational angle from symplectic algorithms increases linearly 
in a staircase-like manner with time. In contrast, the rotational angle of the 
Runge-Kutta algorithm shows a quadratic increase over long times, such as a 
few thousand periods. This result is easy to understand if the rotational angle 
is related to some integral of the energy error. This quadratic increase in
the rotation angle of the LRL vector clearly mirrors the quadratic increase in
phase error of the Runge-Kutta algorithm, as discussed by  Gladman, 
Duncan and Candy\cite{gladman}.  

\begin{figure}%%%%%%%%%%%%%%%%%%%%%%%%%%%%%%%%%%%%%%%%%%%%%%%%%%%%%%%%%
%\topinsert
\noindent
\vglue 0.2truein
\hbox{
\vbox{\hsize=7truein
\epsfxsize=6truein
\leftline{\epsffile{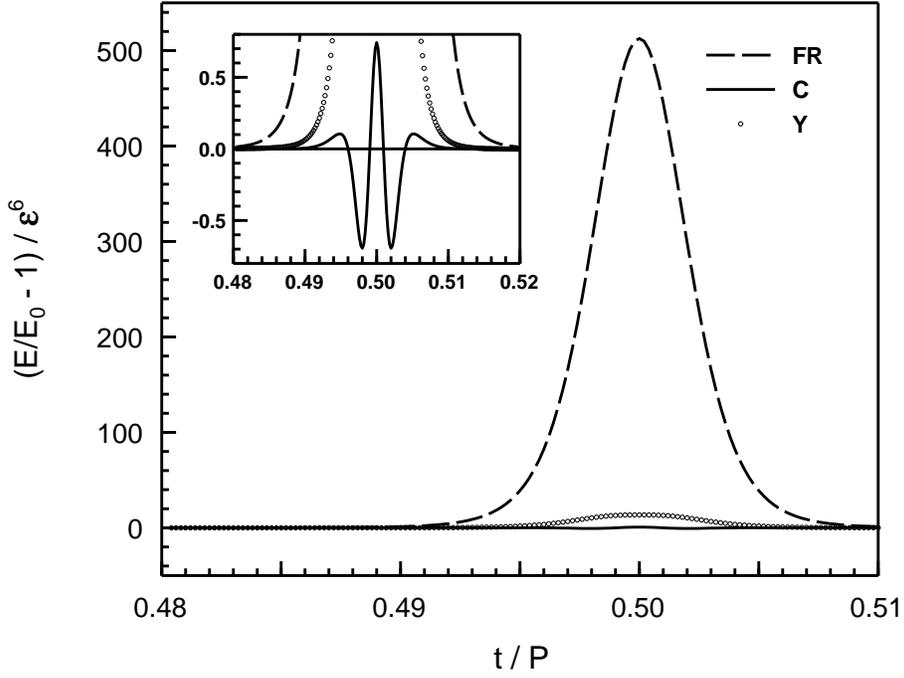}}
}
}
\caption{The normalized energy deviation for 6th order algorithms.
RF and C denote 6th order algorithms generated by a triplet product of 
corresponding 4th algorithms in Fig.\ref{fone}. The insert makes visible the 
energy devivation of algorithm C, which is not visible in the bigger graph. 
The maximum deviations for algorithms
FR, Y and C are 513, 13.6, and 0.74 respectively.
}
\label{fthree}
\end{figure}

\begin{figure}%%%%%%%%%%%%%%%%%%%%%%%%%%%%%%%%%%%%%%%%%%%%%%%%%%%%%%%%%
%\topinsert
\noindent
\vglue 0.2truein
\hbox{
\vbox{\hsize=7truein
\epsfxsize=6truein
\leftline{\epsffile{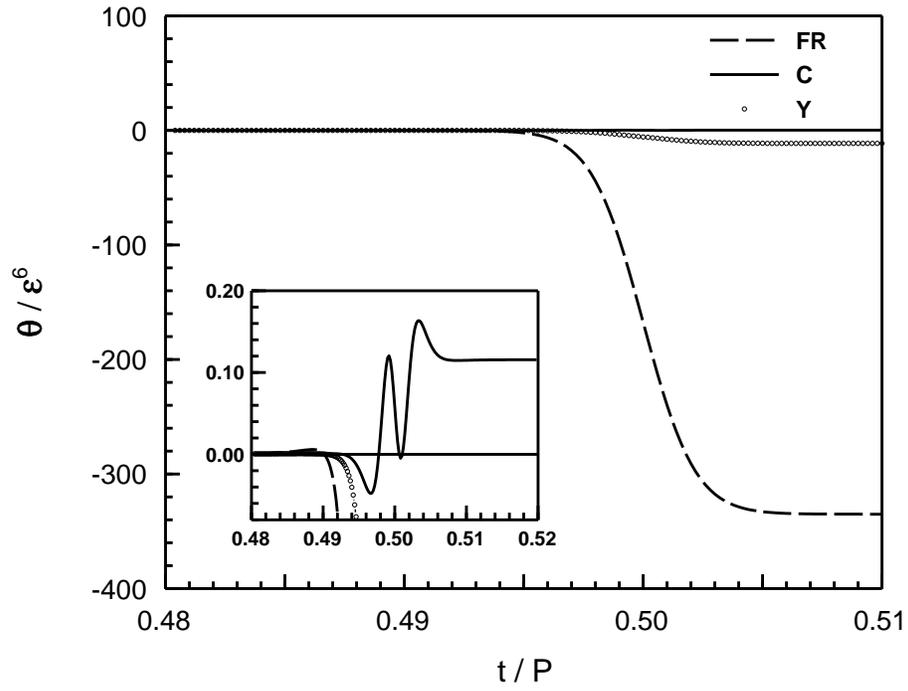}}
}
}
\caption{The step-size independent error coefficient of the rotation
angle of the LRL vector for 6th order algorithms as described in
Fig.\ref{fthree}. The insert makes visible the minute rotation coefficient 
produced by algorithm C. 
}
\label{ffour}
\end{figure}

Fig.\ref{fthree} and Fig.\ref{ffour} show the results when both the 
Forest-Ruth and Chin's C algorithms are iterated to 6th order by the
triplet product construction. Inserts in both detail 
algorithm C's intricate structure. As an added comparison we also 
included results for Yoshida's\cite{yoshid} 6th order algorithm A, which is a 
product of 7 second order algorithms (\ref{altbs}), some with negative 
time steps. 
For Yoshida's algorithm, Forest-Ruth and Chin's C algorithm in 
6th order, the magnitudes of the rotation cofficients after one period are 
11.44, 335.1, and 0.1156 respectively. Yoshida's algorithm is a factor
of 30 better than FR, but algorithm C is a factor of 3000 better. 
Note that if the energy error function
is related to the differential of the the angle error function, the zeros of
the former would correspond to the extrema of the latter. The four 
zero crossings of algorithm C's energy error function are clearly reflected
in the two maxima and two minima of the corresponding angle error function.

\begin{figure}%%%%%%%%%%%%%%%%%%%%%%%%%%%%%%%%%%%%%%%%%%%%%%%%%%%%%%%%%
%\topinsert
\noindent
\vglue 0.2truein
\hbox{
\vbox{\hsize=7truein
\epsfxsize=6truein
\leftline{\epsffile{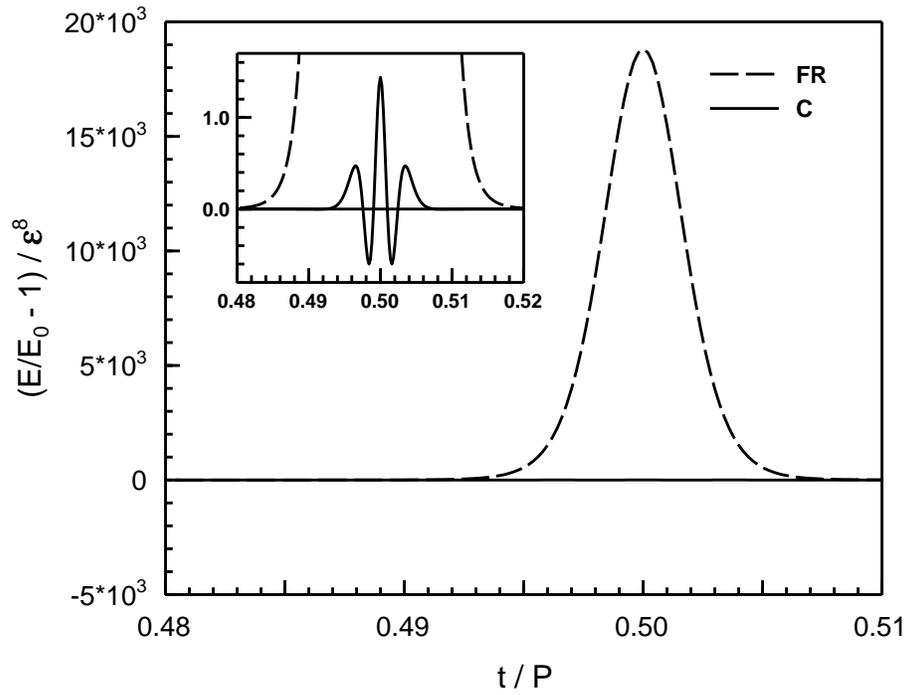}}
}
}
\caption{The normalized energy deviations for 8th order algorithms, as
generated by a triplet product of 6th order algorithm described in
Fig.\ref{fthree}. The insert makes visible the minute energy deviation of
algorithm C. 
}
\label{ffive}
\end{figure}

\begin{figure}%%%%%%%%%%%%%%%%%%%%%%%%%%%%%%%%%%%%%%%%%%%%%%%%%%%%%%%%%
%\topinsert
\noindent
\vglue 0.2truein
\hbox{
\vbox{\hsize=7truein
\epsfxsize=6truein
\leftline{\epsffile{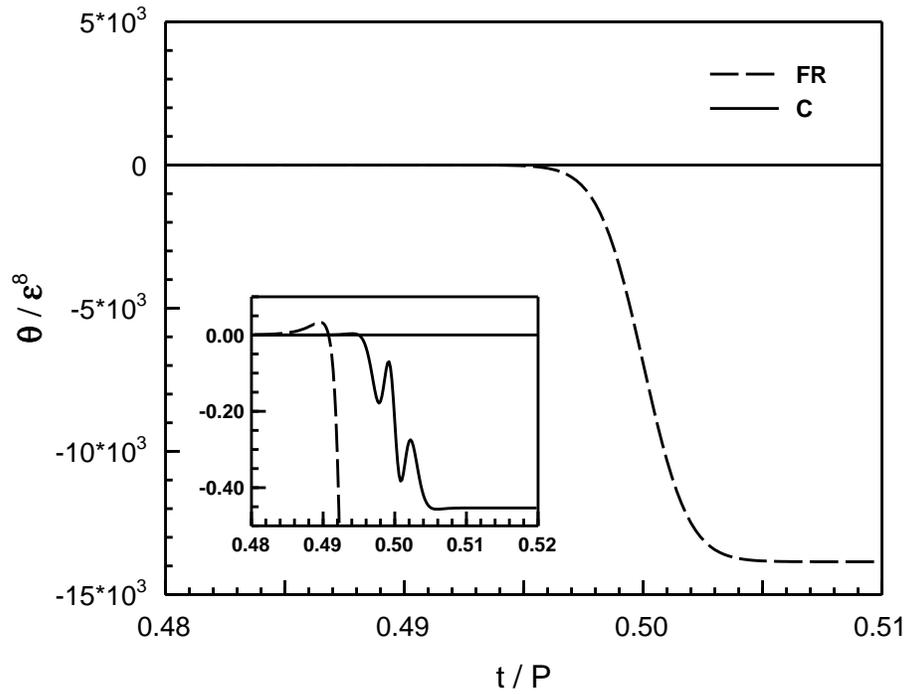}}
}
}
\caption{The step-size independent error coefficient of the rotation
angle of the LRL vector for 8th order algorithms as described in 
Fig.\ref{ffive}. In this order, the algorithm C based algorithm begins
to rotate in the same sense as the FR base algorithm.  
}
\label{fsix}
\end{figure}

\begin{figure}%%%%%%%%%%%%%%%%%%%%%%%%%%%%%%%%%%%%%%%%%%%%%%%%%%%%%%%%%
%\topinsert
\noindent
\vglue 0.2truein
\hbox{
\vbox{\hsize=7truein
\epsfxsize=6truein
\leftline{\epsffile{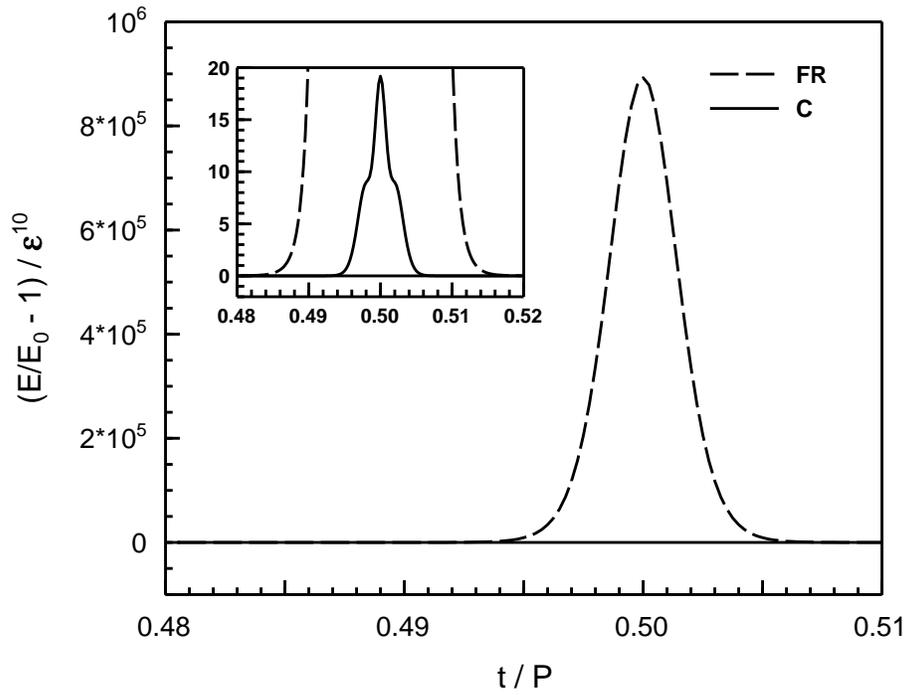}}
}
}
\caption{The normalized energy deviations for 10th order algorithms, as
generated by a triplet product of 8th order algorithm described in
Fig.\ref{ffive}. The insert shows that the characteristic oscillations of
algorithm C are beginning to disappear.
}
\label{fseven}
\end{figure}

\begin{figure}%%%%%%%%%%%%%%%%%%%%%%%%%%%%%%%%%%%%%%%%%%%%%%%%%%%%%%%%%
%\topinsert
\noindent
\vglue 0.2truein
\hbox{
\vbox{\hsize=7truein
\epsfxsize=6truein
\leftline{\epsffile{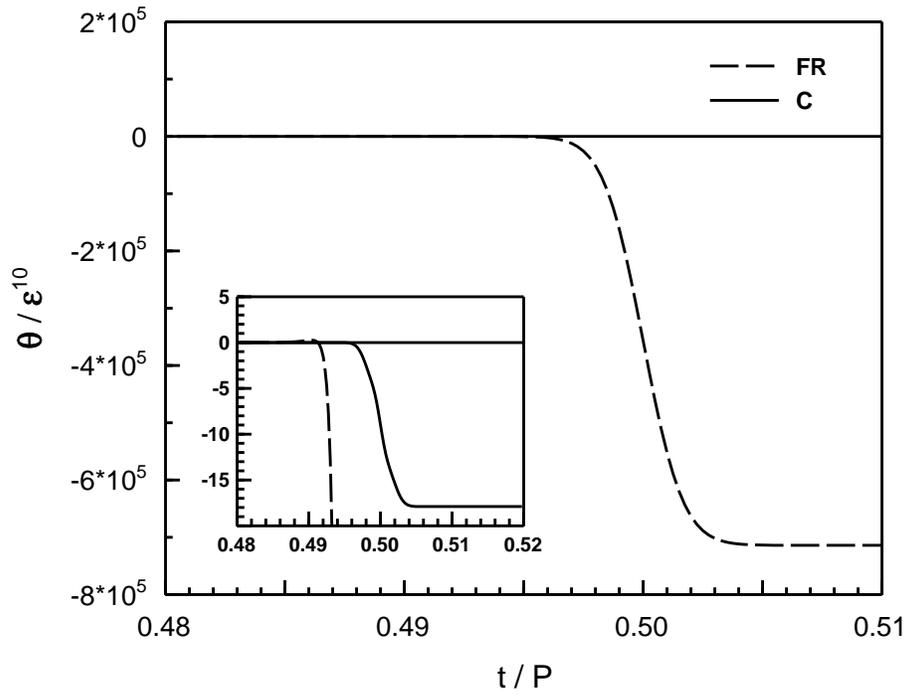}}
}
}
\caption{The step-size independent error coefficient of the rotation
angle of the LRL vector for 10th order algorithms as described in 
Fig.\ref{fseven}. The insert shows that the error coefficient of 
algorithm C begins to look like that of the FR algorithm.
}
\label{feight}
\end{figure}

\begin{figure}%%%%%%%%%%%%%%%%%%%%%%%%%%%%%%%%%%%%%%%%%%%%%%%%%%%%%%%%%
%\topinsert
\noindent
\vglue 0.2truein
\hbox{
\vbox{\hsize=7truein
\epsfxsize=6truein
\leftline{\epsffile{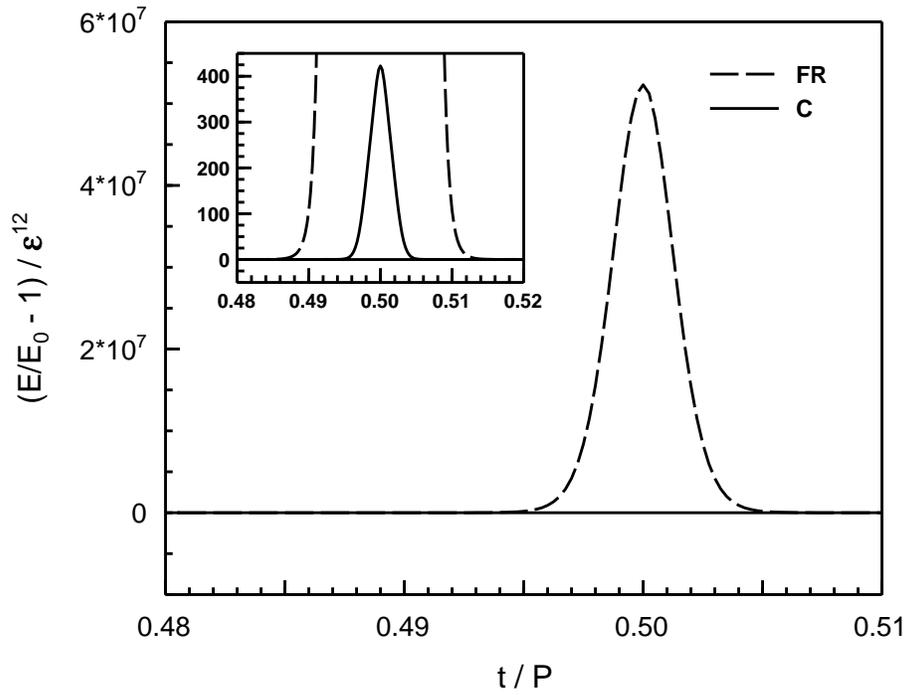}}
}
}
\caption{The normalized energy deviations for 12th order algorithms, as
generated by a triplet product of 10th order algorithm described in
Fig.\ref{fseven}. The insert shows that there is no longer any distinctive 
structure produced by algorithm C.
}
\label{fnine}
\end{figure}

\begin{figure}%%%%%%%%%%%%%%%%%%%%%%%%%%%%%%%%%%%%%%%%%%%%%%%%%%%%%%%%%
%\topinsert
\noindent
\vglue 0.2truein
\hbox{
\vbox{\hsize=7truein
\epsfxsize=6truein
\leftline{\epsffile{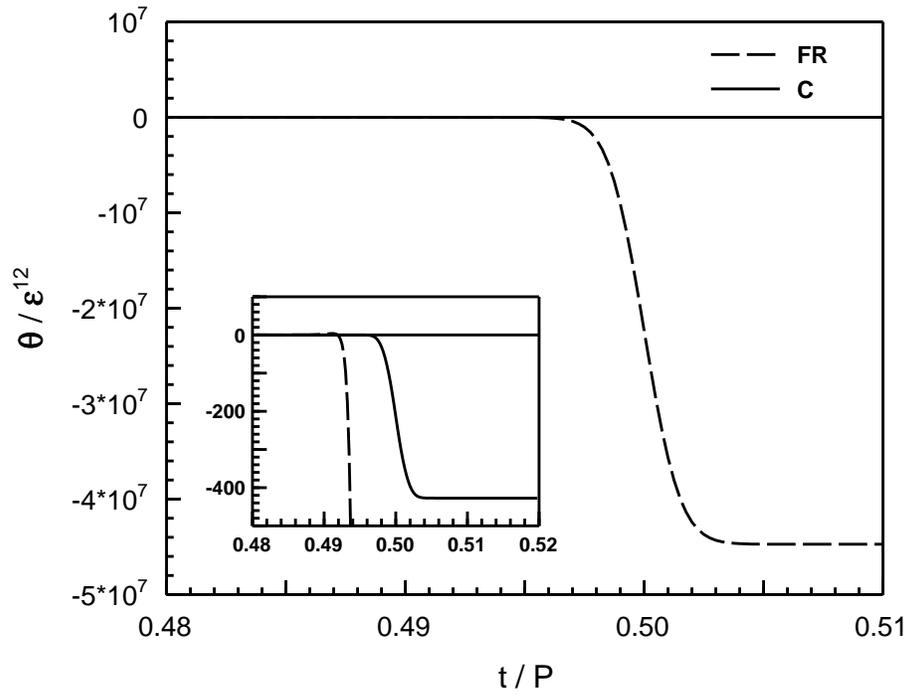}}
}
}
\caption{The step-size independent error coefficient of the rotation
angle of the LRL vector for 12th order algorithms as described in 
Fig.\ref{fnine}. The insert shows that both algorithms have converged to
a similar step-like error function.
}
\label{ften}
\end{figure}

Fig.\ref{ffive} and Fig.\ref{fsix} give results for the 8th order 
iterated algorithms based on the Forest-Ruth and Chin's C algorithm. 
The magnitudes of the angle error coefficients are $1.386\times 10^4$ 
and 0.4532 respectively, giving a ratio of approximately $3\times 10^4$. Algorithm C 
retains its characteristic shape in both the energy and the angle error function. 
Fig.\ref{fseven} and Fig.\ref{feight} give the corresponding results for the
iterated 10th order algorithms. Here the intricate structure in the C algorithm
is beginning to be washed out. At this high order, quadruple numeric precision 
is necessary to extract these coefficient functions smoothly.
The magnitudes of the angle error coefficients are now 
$7.141\times 10^5$ and $17.89$ respectively, giving a ratio of $4\times 10^4$.
Fig.\ref{fnine} and Fig.\ref{ften} give similar results for the 12th order algorithms.
At this point all structures in the C algorithm are gone.     
The magnitudes of the angle error coefficients are now 
$4.473\times 10^7$ and $427.5$ respectively, giving a ratio of $1.05\times 10^5$.

The iteration of algorithms A and B 
of Chin\cite{chin} also produced results that are better than FR based 
algorithms. However, we do not detail their results here because
they are at least one or two orders of magnitude inferior to algorithm C.

\section {Conculsions}

In this work we have shown that higher order force gradient symplectic 
algorithms appear to be superior to non-gradient symplectic alogorithms 
as measured by eneregy conservation and the rotation of the LRL vector. 
While it has been shown earlier that 4th order force gradient algorithms have 
smaller energy error coefficients\cite{chin}, it was not known that this 
advantage would mulitply dramatically when algorithms are iterated to higher 
orders. The conclusion that one should draw may not be that force gradient 
algorithms are better, but that higher order non-gradient algorithms are far 
from optimal. Secondly, we suggested that the rotation of the LRL vector gives 
an intergrated measure of an algorithm's merit when tested on the Kepler problem.

The high accuracy of this class of algorithms seemed ideal for long time integration
of few-body problems, such as that of the solar system\cite{wisdom}. For such 
few-body problems, the evaluation of the force gradient is not excessively difficult.
It would be useful to examine the merit of this class of algorithms in more
physical applications. The distinct 
advantage uncovered in this work, that it is better to iterate a 4th order 
algorithm with all positive time steps, gives further impetus to 
search for an all positive time step 6th order symplectic algorithm.

\acknowledgements
This research was funded, in part, by the U. S. National Science Foundation 
grants PHY-9512428, PHY-9870054 and DMR-9509743.

%%%%%%%%%%%%%%%%%%%%%%%%% Bibliography %%%%%%%%%%%%%%%%%%%%%%%%%%%%%%
%%

\ifpreprintsty\newpage\fi
%%%%% BEGIN Figures Captions%%%%%%%%%%%%%%%%%%%%%%%%
\ifpreprintsty\newpage\fi
%%%%%%%%%%%%%%%%%%%%%  END figures %%%%%%%%%%%%%%%%%%%%%%%%%%%%%%%%%%
%%%%%%%%%%%%%%%%%%%%%  END figures %%%%%%%%%%%%%%%%%%%%%%%%%%%%%%%%%%
%%%%%%%%%%%%%%%%%%%%%  END figures %%%%%%%%%%%%%%%%%%%%%%%%%%%%%%%%%%
%%%%%%%%%%%%%%%%%%%%%  END figures %%%%%%%%%%%%%%%%%%%%%%%%%%%%%%%%%%
%%%%%%%%%%%%%%%%%%%%%  END figures %%%%%%%%%%%%%%%%%%%%%%%%%%%%%%%%%%
%%%%%%%%%%%%%%%%%%%%%  END figures %%%%%%%%%%%%%%%%%%%%%%%%%%%%%%%%%%
%%%%%%%%%%%%%%%%%%%%%  END figures %%%%%%%%%%%%%%%%%%%%%%%%%%%%%%%%%%
%%%%%%%%%%%%%%%%%%%%%  END figures %%%%%%%%%%%%%%%%%%%%%%%%%%%%%%%%%%
%%%%%%%%%%%%%%%%%%%%%  END figures %%%%%%%%%%%%%%%%%%%%%%%%%%%%%%%%%%
%%%%%%%%%%%%%%%%%%%%%  END figures %%%%%%%%%%%%%%%%%%%%%%%%%%%%%%%%%%
\end{document}